\documentclass[twocolumn,twoside,slac_two]{revtex4}
\usepackage{graphicx}
\usepackage{fancyhdr}
\begin{document}

\title{BEPCII/BESIII Upgrade and the Prospective Physics}

%

\author{Weiguo Li}
\affiliation{Institute of High Energy Physics, 19 Yuquan Road, Beijing 100049, China}

\begin{abstract}
The Status of BEPCII/BESIII upgrade is described, BEPCII is
designed to reach a luminosity of $10^{33}/cm^{-2}s^{-1}$ at c. m.
energy of 3.89GeV, and BESIII detector is fully rebuilt. The
project is on track and is progressing well. The machine and
detector are expected to commission together at the late half of
2007, and start to take some engineer run end of 2007 or beginning
of 2008.
\end{abstract}

\maketitle



\section{Introduction}
Beijing Electron Positron Collider (BEPC) started data taking in
1989, and they were upgraded in 1996, and the upgraded detector is
called BESII~\cite{besiidetector}. BES had taken data until the
April of 2004, many results have been obtained using the data
samples collected at J/$\psi$, $\psi(2S)$, $\psi(3770)$ and the
scan data from 2.0 to 5.0 GeV.

After more 15 years data taking, the machine and detector are no
more competitive to produce first grade physics at this energy
region, especially after CESR reduced its energy to the
$\tau$-charm energy region. After several years of preparation,
the Chinese Government gave a green light to begin the
BEPCII/BESIII upgrade at the end of 2003.

  BEPCII is a two rings machine built in the existing tunnel, electrons
and positrons are circulated in separate rings and only collider
in the interaction point, multi-bunches and micro-beta are used to
increase the machine luminosity. Super-conducting cavities and
super-conducting quadruples are used. The Linac has to be upgraded
also to increase its energy and currents, especially to meet the
needs to inject the positions to the ring in a reasonable short
time.

  The detector has to be totally rebuilt, with a 43 layers small-cell
main drift chamber (MDC), a time of flight (TOF) system, and an
electro-magnetic calorimeter Made of CsI, a super-conducting
magnet with a field of 1 tesla, and resistive plate chambers (RPC)
are inserted in the magnet yoke to serve as the muon counters.

  The whole project is scheduled to complete in 2008, with a goal that
the luminosity of the machine should reach $3\times 10^{32}cm^{-2} s^{-1}$ at the end of 2008.

\section{BEPCII}
The main parameters of machine parameters are listed in Table~I.

\begin{table}[ht]
\begin{center}
\caption{BEPCII Design Goal.}
\begin{tabular}{|c|c|}
\hline Energy Range & 1-2.1 GeV
\\
\hline Optimum energy & 1.89 GeV
\\
\hline Luminosity & $1\times 10^{33}cm^{-2}s^{-1}$ at 1.89 GeV\\
\hline Injection & Full energy inject.: 1.55-1.89 GeV\\
 & Position Injection $>$ 50 mA/min
\\
\hline Synchrotron mode & 250 mA at 2.5 GeV\\
\hline
\end{tabular}
\end{center}
\end{table}

\begin{table}[ht]
\begin{center}
\caption{BEPCII Linac Achieved Performances.}
\begin{tabular}{|c|c|c|c|}
\hline Parameters &  Design  & Achieved & BEPC
\\
\hline Beam energy (GeV) & 1.89 & 1.89($e^-$);1.89($e^+$) & 1.55
\\
\hline $e^+$ Current  & 40 & $>$ 63 & 4\\
\hline $e^-$ Current  & 500 & $>$ 500 & 50\\
\hline Repetition rate (Hz) & 50 & 25-50 & 25
\\
\hline $e^+$ emit. (mm-mrad)  & 1.60 & $>$ 0.93 (1.89 GeV) & 1.70\\
\hline $e^-$ emit. (mm-mrad) & 0.20 & $>$ 0.30 (1.89 GeV) & 0.58\\
\hline $e^+$ energy spread (\%)  & $\pm0.5$ & $\pm$ 0.50 (1.89 GeV) & $\pm$ 0.8\\
\hline $e^-$ energy spread (\%) & $\pm0.5$ & $\pm$ 0.55 (1.89 GeV) & $\pm$ 0.8\\
\hline
\end{tabular}
\end{center}
\end{table}

  To increase the injection energy and current to the rings, the Linac
has been upgrades, the main systems are new acceleration tubes,
new klystrons and modulators, new positron source, new electron
gun and increase the beam repetition rate to 50 Hz, modified
vacuum system, and other relevant modifications. The Linac has
been debugged and tuned, some of the performances have already
reached the designed values, and now the main efforts are to
understand the operation of the Linac and to make the running
condition to be stable and install the phase control system. The
achieved performances of the Linac are listed in Table II.

\begin{table*}[t]
\begin{center}
\caption{BEPCII Main Parameters.}
\begin{tabular}{|c|c|c|c|}
\hline Parameters &  Unit & BEPCII & BEPC
\\
\hline Operation energy & GeV & 1.0-2.1 & 1.0-2.5
\\
\hline Injection energy($E_{inj}$) & GeV & 1.55-1.89 & 1.3
\\
\hline Circumference(C) & m & 237.5 & 240.4
\\
\hline $\beta^*$-function at IP($\beta^*_x$/$\beta^*_y$) & cm &
100/1.5 & 120/5
\\
\hline Tunes($\nu_x/\nu_y/\nu_s$) & & 6.57/7.61/0.034 &
5.8/6.7/0.02
\\
\hline Hor. natural emittance($\theta_{x0}$) & mm-mr & 0.14 at 1.89 GeV & 0.39 at 1.89 GeV \\
\hline Dampling time ($\tau_{x}/\tau_{y}/\tau_{z}$) &  &
25/25/12.5 at 1.89 GeV & 28/28/14 at 1.89 GeV
\\
\hline RF frequency ($f_{rf}$) & MHZ & 499.8 & 199.533
\\
\hline RF voltage per ring ($V_{rf}$) & MV & 1.5 & 0.6-1.6
\\
\hline Number of bunches &   & 93 &2$\times$1\\
\hline Bunch spacing & m & 2.4  & 240.4
\\
\hline Beam current colliding & mA & 910 \@ 1.89GeV & ~2$\times$35
\@ 1.89 GeV
\\
\hline Bunch length($\sigma_l$) & cm  & ~1.5 &~5\\
\hline Impedance $|Z/n|_0$ & $\Omega$ & ~0.2 & ~4
\\
\hline Crossing angle & mrad & $\pm 11$   & 0
\\
\hline Beam lifetime & hrs. & 2.7 & 6-8
\\
\hline Luminosity\@1.89 GeV & $10^{31}
cm^{-2}s^{-1}$ & 100   & 1
\\
\hline
\end{tabular}
\end{center}
\end{table*}

\begin{figure*}[t]
\centering
\includegraphics[width=135mm]{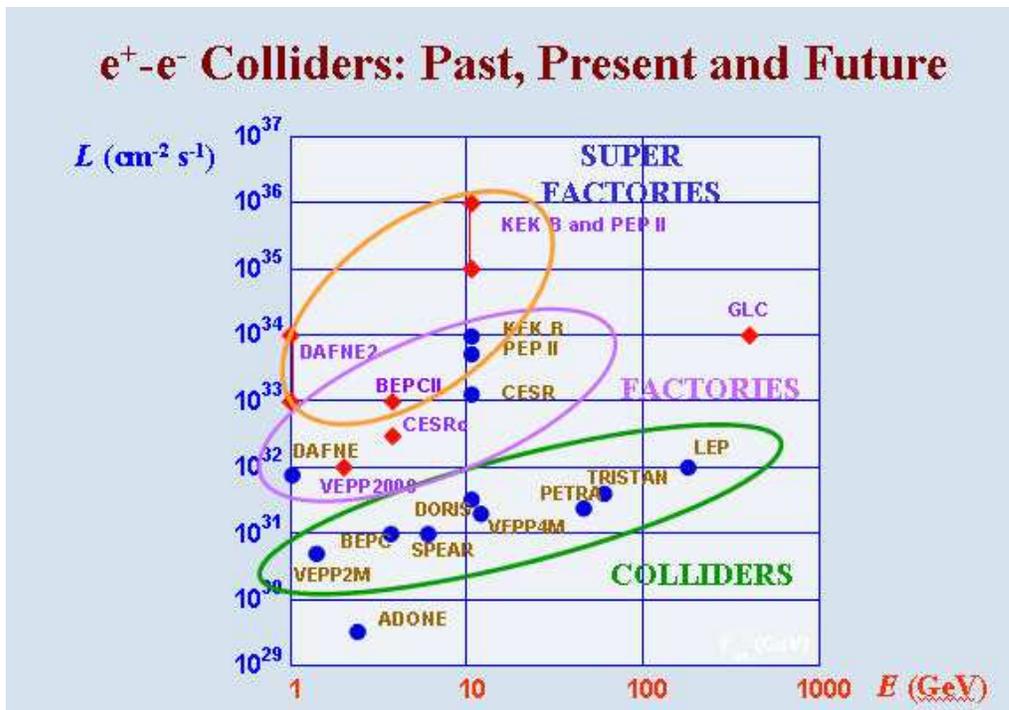}
\caption{Performances of $e^+e^-$ colliders.}
\end{figure*}

The main parameters of the storage rings are listed in Table III.
BEPCII storage rings have the following main systems: the
super-conducting rf cavities and their power supplies and control
system; the beam pipes; the magnets and their power supplies; the
kickers; the beam instrumentations; the vacuum system; the control
system. The dipole magnets of old ring are modified to be put in
the outer ring. Up till now, most of the hardware devices are
delivered, and the pieces not on site yet are scheduled to be
delivered on time. The current plan is to assemble all the ring
hardware pieces except the super-conducting quadruple magnets, for
which more time is needed to get them tested and to measure the
field with detector magnet together. The plan is to commission the
whole rings without the quadruples first and provide some beam
time for synchrotron use, then the quadruples are to be moved in
the beam line to be tested together with other systems, after
certain performance goals are met, especially certain amount of
current should be reached so to verify the large currents can be
run with certain luminosity and reasonable backgrounds to the
detector, before the detector moves in to be debugged together
with the machine. The machine and detector are expected to be
commissioned together at the late half of 2007, and start to take
some engineer run end of 2007 or beginning of 2008. The
performance of BEPCII will be among the 2nd generation of $e^+e^-$
colliders, as shown in Figure 1.

\section{BESIII Detector}

  The BESIII detector~\cite{besiiidetector} is a completely new detector. Figure 2 shows
the schematic of BESIII detector.

\begin{figure*}[t]
\centering
\includegraphics[width=135mm]{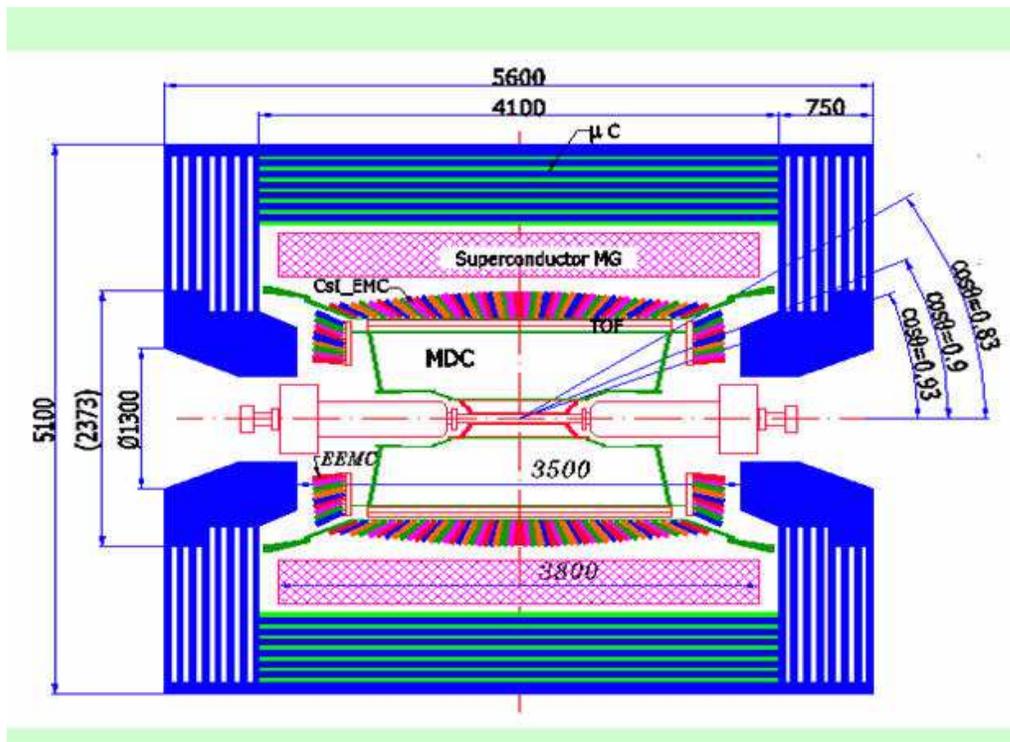}
\caption{Schematic of BESIII detector.}
\end{figure*}

The beam pipe is made of two layers of Be pipes, with the
thickness of 0.8 mm and 0.5 mm respectively, and a cooler is
passed through between the pipes to take out the heat caused by
the beams. MDC is the most inside detector component, it has 43
sense wires from inside to outside. To make room for the machine
super-conducting magnet, the endplates are made from a cone shaped
central section and a slightly inclined outer plate. The endplates
are manufactured with the average hole accuracy better than 25
$\mu m$. Small cell structure is used with a half width of cell to
be 6 mm for inner layers, and 8.1 mm for outer layers. The sense
wire is made of 25 $\mu m$ tungsten, the field wire is made of 110
$\mu m$ gold plated Al. Totally there are about 28K wires. The gas
chosen is $He(60)/C_3H_8(40)$ used by CLEOIII. All the wires have
been strung and the tension and the leakage current of all wires
are measured and satisfy the requirements. The electrons readout
system adopts CERN HPTDC to measure the drift time, the prototype
electron system is tested, and the performances meet the design
requirements. A small chamber was tested in the KEK beam line, and
the space resolution and dE/dx resolution are satisfactory. The
expected chamber performances are, wire resolution to be better
than 130 $\mu m$, and the momentum resolution is expected to
better than 0.5\% for 1 GeV tracks, two contributions from wire
position resolution and multiple scattering are:$\sigma_{pt}/pt$ =
$0.32\%/p$ and $ 0.37\%/\beta$ respectively. The outmost sense
wire covers a polar angle of $cos(\theta)$ = 0.83, and the maximum
acceptance is $cos(\theta)$ = 0.93, roughly reach 20th sense wire
layer. The expected dE/dx resolution is about $6\%$.

  Mounted on the MDC are two layers of TOF counters, each layer has 88 counters
with a thickness of 5 cm and the two layers are staggered by half
a counter to make a full coverage in the polar angle of
$cos(\theta)$ = 0.82. There are photo-tubes at both ends of each
counter to read the signals. And there are 48 single-layer
counters in each side of the endcap region, and the signal is read
out only at the inner end. The counters are mounted on the endcap
EM calorimeter. BC408 and BC404 scintillators by Bicro are chosen
for barrel and endcap respectively. The photo-tubes used are
R2490-50 by Hamamatsu. From the test beam, 90 and 100 ps time
resolutions are achieved for barrel and endcap respectively, as
expected in the design.

  Outside of TOF system is a CsI crystal EM calorimeter. The typical
crystal has a dimension of $5\times 5 cm^2$ in the front face and
$6.5\times 6.5 cm^2$ in the rear face, the length is 28 cm,
corresponding to 15 radiation length. There are 5280 crystals in
barrel and 960 in the endcap, 480 at each side.  Two photo diodes
(Hamamatsu S2744-08) with a photosensitive area of 10 mm $\times$
20 mm are mounted on the rear face of each crystal to read the
lights out. The prototype for readout electronics is tested to
have a noise level of 1000 electrons. The expected energy
resolution will be $2.5\%$ for 1 GeV photons. The crystals are
made by Crismatec(France), Shanghai Institute of Ceramic and
Beijing Hamamatsu. Most of the crystals are delivered and they met
the specifications of dimensions, light yield and the radiation
hardness.
  The mechanical structure of the calorimeter is designed such that there
are no walls between crystals to reduce the dead material, the
crystals are fixed to a support structure by 4 screws. $N_2$ will
be flew in the crystal container to maintain the humidity inside,
and the front-end electronics will be cooled by water. The EM
calorimeter is scheduled to be  assembled around the end of 2006.

  Outside of EM calorimeter is the super-conducting magnet, its designed field is
1.0 tesla. The magnet uses the inner winding technique with the
coil wound inside a Al support cylinder which in turn cooled by
liquid He circled in the pipe welded on the outer surface of the
support cylinder, the Al stabilized NbTi/Cu
 coil is made by Hitachi company, with the nominal operating current of about 3700 A, which
has a stored energy of about 10 MJ. The field in the MDC volume
has an uniformity of better than 5$\%$, the field will be measured
with an accuracy of better than 0.25$\%$. The magnet is in the
stage of testing the value box and after that to install the valve
box with the magnet coil assembling, then the cryogenic system
will be connected to the magnet and the whole magnet will be
cooled and tested.

  The return yoke is also served as the absorber of the muon detectors. The
active muon detector is the Resistive Plate Chamber (RPC), there
are 9 layers muon chambers in the barrel and 8 layers in the
endcap, with two orthogonal strip readout alternatively layer by
layer. The muon chambers are inserted in the steel slots in the
yoke. The special feature of the RPC made in China is that there
is no linseed oil used in the gap of RPC. The gas used is
$Ar:C_2H_2F_4:Iso-Butane = 50:42:8$. All the RPC are installed in
the yoke and the efficiencies and the dark currents measured are
quite good, to the same level as those RPC with linseed oil used
in other experiments.

  The whole detector weights about 800 tons. And the detector hall will
be air-conditioned, to control the temperature at $22\pm 2^oC$,
and the humidity below $55\%$.

  The electronics and trigger use pipelined arrangement, with a trigger latency
of 6.4 $\mu$s. The trigger will use TOF, MDC and Muon information
to make decisions. The simulation shows that good (almost 100$\%$)
efficiency and a good background rejection can be achieved. The
maximum trigger rate will be at J/$\psi$ energy with a total
trigger rate of about 4000 Hz. At this energy, the good event rate
will be about 2000 Hz. The DAQ bandwidth will be about 50 Mbytes
per second. The readout system is based on VME. The offline
analyzes package is under development, the preliminary version for
event simulation and reconstruction is ready, the calibration
codes and physics analyzes code are being worked on. The offline
system will go through another two releases to get the package
tested before the real data are taken.

  The main designed detector performances are listed in Table IV,
a comparison with CLEOc is made. Hopefully, some of the parameters
may be better than designed.
\begin{table}[h]
\begin{center}
\caption{BESIII Performances Compared with CLEOc.}
\begin{tabular}{|c|c|c|}
\hline Detector &  BESIII & CLEOc
\\
\hline   & $\sigma_{xy}$ = 130$\mu$m & 90$\mu$m
\\
 MDC  & $\Delta p/p$(1 GeV) = 0.5$\%$ & 0.5$\%$\\
       & $\sigma_{dE/dx}$ = 6-7$\%$  &  6$\%$\\
\hline EMC  & $\Delta E/\sqrt{E}$(1 GeV) = 2.5$\%$  & 2.2$\%$
\\
    & $\sigma_z$ = 0.6cm/$\sqrt{E}$ & 0.5cm/$\sqrt{E}$ \\
\hline TOF & $\sigma_T$ = 100-110ps & Rich\\
\hline $\mu$ counter & 9 layers & --- \\
\hline magnet & 1.0 tesla & 1.0 tesla \\
\hline
\end{tabular}
\end{center}
\end{table}

\section{Prospective Physics}

  The rich physics topics at this energy region will require BESIII to take the data
at J/$\psi$, $\psi(2S)$, $\psi(3770)$, and some energy point for
$D_S$, also data at $\tau$ threshold and some scan data to measure
hadronic cross-section in this energy region will be collected.
The yearly yield of events are listed in Table V for different
energy points the data are to be taken.
\begin{table*}[t]
\begin{center}
\caption{BESIII Yearly Event Production.}
\begin{tabular}{|c|c|c|c|c|}
\hline Resonance &  Energy(GeV) & Peak Lum. & Physics Cross & Nevents/yr
\\
    & & $10^{33}cm^{-2}s^{-1}$ & Section(nb) &
\\
\hline J/$\psi$ & 3.097 & 0.6 & 3400 & $10\times 10^9$
\\
\hline $\tau$ & 3.670 & 1.0 & 2.4 & $12\times 10^6$
\\
\hline $\psi(2S)$ & 3.686 & 1.0 & 640 & $3.2\times 10^9$
\\
\hline $D^0\bar{D^0}$ & 3.770 & 1.0 & 3.6 & $18\times 10^6$
\\
\hline $D^+D^-$ & 3.770 & 1.0 & 2.8 & $14\times 10^6$ \\
\hline $D_SD_S$ & 4.030 & 0.6 & 0.32 & $1.0\times 10^6$
\\
\hline $ D_SD_S$ & 4.140 & 0.6 & 0.67 & $2.0\times 10^6$
\\
\hline
\end{tabular}
\end{center}
\end{table*}

  The new X(1835) and other near threshold enhancements recently observed at
BES will be studied with 100 times more data, X(3872) and states
observed at 3940 MeV and 4260 MeV in other experiments may be
studied in details. There are more scalars existed in this energy
region to be accommodated in naive quark model, these states will
be thoroughly studied.

  The $\eta_C$, $\chi_{CJ}$ and $h_C$ can be studies with large statistics. The
$\rho\pi$ puzzle will be studied with more decay channels and with
better accuracy and different models can be tested and developed
to explain the mechanism behind that.

  A data sample at $\psi(3770)$  are to be taken, it will enable the
measurements related with D decays to reach a new precision, for
example, the $V_{cd}$, $V_{cs}$ can be measured to a statistic
accuracy of about $1.6\%$, with a data sample of total accumulated
luminosity of 20 $f_b^{-1}$. And the $D^0\bar{D^0}$ mixing and CP
violation will be searched and studied.
  With huge data samples, the systematic errors should be well understood,
and a lot of analyzes will adopt partial wave analyzes (PWA) to
fully understand the decay dynamics. Efforts are being made and
will be strengthened.

Right now, BES Collaboration has about 18 Chinese institutes and
universities, and physicists from United State, Japan, Germany,
Sweden and Russia have joined. BES welcomes new collaborators to
join this exciting research project, which should last for at
least 10 years. A conference called Charm2006 will be held in
Beijing in June this year to discuss the physics potentials at
BESIII and a US-CHINA workshop on the BESIII collaboration will be
held right afterwards. All interested persons are welcome.

\bigskip 

\end{document}